\documentclass[prb,aps,twocolumn,floatfix,amsmath,amssymb,tightenlines, showpacs, superscriptaddress]{revtex4}
\usepackage[pdftex]{graphicx}
 \usepackage{amsmath}
\usepackage{flushend}
\usepackage{MnSymbol}
\usepackage[utf8x]{inputenc}
\usepackage[T1]{fontenc}
\usepackage{amssymb}
\usepackage{amsfonts}
\usepackage{bm}
 \usepackage{amsmath} 
\usepackage{subfigure}
\usepackage{lipsum}
\usepackage{amsfonts} 
\usepackage{amssymb, mathrsfs}
\usepackage{braket}
\usepackage{graphicx} 
\usepackage[usenames]{color} 

\usepackage{bbm}
\def\beq{\begin{equation}}
\def\eeq{\end{equation}}
\def\bsp{\begin{split}}
\def\esp{\end{split}}
\def\bea{\begin{eqnarray}}
\def\eea{\end{eqnarray}}
\def\ba{\begin{array}}
\def\ea{\end{array}}

\def\dg{\dagger}

\def\lb{\left(}
\def\rb{\right)}

\def\l.{\left.}
\def\r.{\right.}

\def\ra{\rangle}
\def\la{\langle}

\def\bo{\bold{k}}

\begin{document}

\title{A Large-$S$ Study of Quantum Kagome Ice}
\author{S. A. Owerre}
\email{sowerre@perimeterinstitute.ca}

\affiliation{Perimeter Institute for Theoretical Physics, 31 Caroline St. N., Waterloo, Ontario N2L 2Y5, Canada.}
\affiliation{African Institute for Mathematical Sciences, 6 Melrose Road, Muizenberg, Cape Town 7945, South Africa.}
\author{A. A. Burkov}
\affiliation{Department of Physics and Astronomy, University of Waterloo, Waterloo, Ontario N2L 3G1, Canada.}

\affiliation{ITMO University, Saint Petersburg 197101, Russia. }

\author{Roger G. Melko}
\affiliation{Perimeter Institute for Theoretical Physics, 31 Caroline St. N., Waterloo, Ontario N2L 2Y5, Canada.}
\affiliation{Department of Physics and Astronomy, University of Waterloo, Waterloo, Ontario N2L 3G1, Canada.}

\begin{abstract}
We present a large-$S$ study of a quantum spin ice Hamiltonian, introduced by Huang {\it et~al.}~[\prl {\bf 112}, 167203 (2014)], on the kagome lattice.  This model involves a competition between the frustrating Ising term of classical kagome ice, a Zeeman magnetic field $h$, and a nearest-neighbor transverse spin-flip term $S^x_iS^x_j - S^y_iS^y_j$.  Recent Quantum Monte Carlo (QMC) simulations by Carrasquilla, {\it et~al.,}~[Nature Communications {\bf 6}, 7421 (2015)], uncovered lobes of a disordered phase for large Ising interaction and $h\neq 0$ -- a putative quantum spin liquid phase. Here, we examine the nature of this model using large-$S$  expansion. We show that the ground state properties generally have the same trends with those observed in QMC simulations.  In particular, the large-$S$ ground state phase diagram captures the existence of the disordered lobes.

\end{abstract}


\maketitle

\section{Introduction}
\parskip 0pt 
Quantum Spin Ice (QSI) is the name given to a class of magnetic system on the pyrochlore lattice which exhibit a competition between classical 
frustration and quantum fluctuations.\cite{you,gin, zhi, sun1,sun2,sun3,sun4,sun4a, sun5,sun6,sun7, mat1, uda}  The particular kind of quantum fluctuations acting within the set of classical spin ice states
(characterized by their induction of a Pauling residual entropy) leads to a promising avenue to search for the elusive quantum spin liquid\cite{balent} (QSL)
state among an actively studied set of real compounds.\cite{sun1,sun2,sun3,sun4,sun5,sun6,sun7,fen,kim, sun4a, Huang, balent, mat1, uda} 
Recently, Huang, Chen, and Hermele\cite{Huang} have proposed a specialized model of QSI, relevant to strong spin-orbit coupled
materials exhibiting Kramers doublets with dipolar-octupolar character in $d$-and $f$-electron systems.  The model has several competing diagonal and off-diagonal
terms in the Hamiltonian, and is believed to harbor both gapless and gapped spin liquids in different parameter regimes.
From this three-dimensional (3D) Hamiltonian, the model can be reduced to a two-dimensional (2D) Hamiltonian of interacting spins on the kagome lattice. \cite{uda, mat1,juan,sun6} 
This model, which we call {\it Quantum Kagome Ice} (QKI), opens a theoretical and experimental avenue to search for two-dimensional QSL phases within the broader context of QSI physics.
Most importantly from a theoretical perspective, the quantum kagome ice Hamiltonian (like the original Huang-Chan-Hermele Hamiltonian) 
is devoid of the quantum Monte Carlo (QMC) ``sign problem'', allowing for systematic study of the phase diagram via large scale computer simulations.

  As mentioned above, the model Hamiltonian  for quantum kagome ice studied by  Carrasquillia {\it et  al.,}~\cite{juan}  originates from a certain class of $d$-and $f$-electron systems in which dipolar-octupolar Kramers doublets arise on the sites of the pyrochlore lattice.\cite{Huang} In the localized limit, this model maps to a well-known XYZ Hamiltonian on the pyrochlore lattice. It is known to exhibit two distinct phases on the pyrochlore lattice --- dipolar QSI and octupolar QSI phases. \cite{Huang}   In the presence of a magnetic field along the [111]  crystallographic direction there is a partial lifting of degeneracy.  The magnetic field pins one spin per tetrahedron  and the model effectively decouples into  alternating kagome and triangular layers of the original pyrochlore structure. \cite{uda, mat1,juan,sun6} The resulting  quantum kagome ice Hamiltonian is given by \cite{Huang, juan}
\begin{align}
&H_{XYZ}=\sum_{\la lm\ra}[ \mathcal{J}_zS_{l}^zS_{m}^z-\frac{\mathcal{J}_1}{2}\lb S_{l}^+S_{m}^- + S_{l}^-S_{m}^+\rb\nonumber\\&+\frac{\mathcal{J}_2}{2}\lb S_{l}^+S_{m}^+ + S_{l}^-S_{m}^-\rb]-h\sum_l S_l^z,
\label{k1}
\end{align}
where $S_{\pm}=S_x\pm iS_y$ can be used to transform Eq.~\eqref{k1} into an XYZ Hamiltonian. The above model [Eq.~\eqref{k1}] is regarded as  a projection of the QSI Hamiltonian on a $3$D pyrochlore lattice onto the $2$D kagome lattice Hamiltonian .\cite{Huang, mat1,mat2,mat3,mat4}  
The previously-studied U(1)-invariant XXZ model \cite{xu, isa, roger, kedar, senn} (equivalent to hard-core bosons) is recovered when $\mathcal{J}_2=0$. As $\mathcal{J}_2$ is turned on, the U(1)  symmetry is explicitly broken, and Eq.~\eqref{k1} retains only $Z_2$ symmetry: $S_{l,m}^\pm \to -S_{l,m}^\pm$; $S_{l,m}^z\to S_{l,m}^z$.  Note that the sign of the $\mathcal{J}_2$ term can be changed by a canonical unitary transformation, $S_{l,m}^\pm \to \pm i S_{l,m}^\pm$, that leaves all other terms invariant.  This unitary transformation corresponds to a $\pi/2$ rotation about the $z$-axis.

Recently, it has been shown using non-perturbative, unbiased QMC simulations,\cite{juan} that Eq.~\eqref{k1} with $\mathcal{J}_1=0$  promotes a QSL phase. The Hamiltonian exhibits several distinct phases as a function $ h/\mathcal{J}_z$ and $ \mathcal{J}_2/\mathcal{J}_z$. 
At $\mathcal{J}_1=0$ and small $h$, there is an in-plane canted ferromagnet (CFM), with spins partially ordered along the $x$- or $y$-direction (depending on the sign of $\mathcal{J}_2$), that exists on the entire $ \mathcal{J}_2/\mathcal{J}_z$ axis, arising from a spontaneous breaking of the Hamiltonian $Z_2$ symmetry.
For very large $h$, a fully polarized (FP) phase occurs with spins aligned along the $z$-direction. 
The nontrivial phases appear in the regime $h/\mathcal{J}_z\neq 0$ and $ \mathcal{J}_2/\mathcal{J}_z<0.5$ and are called ``lobes'' on the phase diagram.  The lobes have magnetization per site $m \approx \pm 1/6$ depending on the sign of $h$,
 {arising from the constraint that two spins point up and two down (or vice-versa), per unit cell of 3 sites}.
These lobes were described in Ref.~[\onlinecite{juan}] as a candidate gapped featureless QSL state, with no evidence of conventional ordering of any kind.

{In this paper, we aim to complement the recent quantum Monte Carlo simulation work by Carrasquillia {\it et al.,}~\cite{juan} through an analytical large-$S$ approach.\cite{pw, jon, joli} We re-examine the phases observed in the QMC study from the perspective of the spin-wave expansion.  We begin with $h=0$, where the classical ground state is highly degenerate.  To first-order in $1/S$, we show that quantum fluctuations lift the degeneracy by selecting a unique ground state -- mirroring what is seen in the QMC. }  Next, in the $m \approx \pm 1/6$ lobes ($h\neq 0$ and $\mathcal{J}_z\gg \mathcal{J}_2$ with $\mathcal{J}_1=0$) the linear spin-wave analysis produces a ground-state energy, magnetization, and ``condensate fraction'' with similar features as the QMC,
 including the phase transition out of the CFM phase.
The excitation spectrum at $\mathcal{J}_1=0$ displays no soft modes for a wide range of parameter space.  
Thus, at the level of our large-$S$ theory, there is no indication of any tendency toward any magnetic order inside the putative spin
liquid lobes.

\begin{figure}[ht]
\centering
\includegraphics[width=3in]{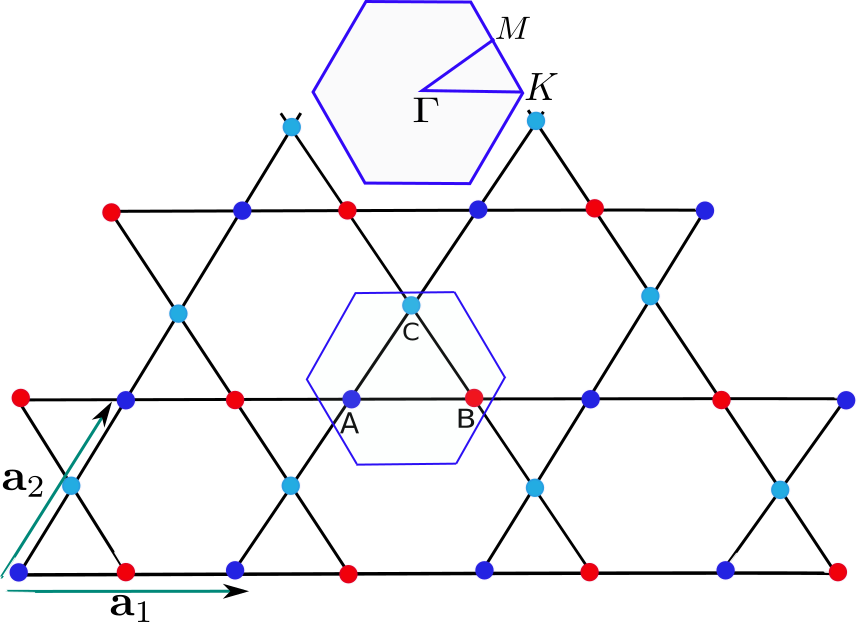}
\caption{Color online. The $\bold q=0$ kagome lattice with three-sublattice unit cell denoted by $A,B,C$. The blue hexagon is the Brillouin zone with $\Gamma= (0,0)$; $K=(2\pi/3, 0)$; $M=(\pi/2, \pi/2\sqrt{3})$. The lattice vectors $\bold{a}_{1/2}$, are given by $\bold{a}_1 =2a\lb 1, 0\rb;\thinspace \bold{a}_2 =2a\lb\frac{1}{2}, \frac{\sqrt 3}{2}\rb, \bold{a}_3=\bold{a}_2-\bold{a}_1$. We set $a=1/2$ in the computations.}
\label{kagome}
\end{figure}

\section{The classical phase diagram} \label{class}
We begin with a study of the classical ground states of the Hamiltonian in Eq.~\eqref{k1}. A crucial observation (made above) is that the sign of the $\mathcal{J}_2$ term is not unique, i.e., ~it can be changed by a canonical unitary transformation. 
At zero magnetic field, and in the limit $\mathcal{J}_1=0$; $\mathcal{J}_2\gg \mathcal{J}_z$, the ground state of Eq.~\eqref{k1} is an in-plane CFM state ordered along the $x$- or $y$-direction depending on the sign of $\mathcal{J}_2$. In contrast, the limit $\mathcal{J}_1=\mathcal{J}_2=0$ exhibits extensive accidental degeneracy; it corresponds to the classical spin ice on kagome lattice.\cite{gin,zhi,you}  In the large-$S$ limit, the spin operators in Eq.~\eqref{k1}  can be represented as classical vectors\cite{sow} which can be written as
 $\bold{S}_l= S\bold{n}_l$, where $\bold{n}_l=\lb\sin\theta_l\cos\phi_l, \sin\theta_l\sin\phi_l,\cos\theta_l \rb$
 is a unit vector.
The classical Hamiltonian under the unitary transformation, $S_{l,m}^\pm \to \pm i S_{l,m}^\pm$ is given by,
\begin{align}
\mathcal{E}_c/{N}S^2&= -\sum_\mu h_\mu+  \sum_{\mu\nu}\xi_{\mu\nu} \big[ \mathcal{J}_z\cos\theta_\mu\cos\theta_\nu \nonumber \\
& -\mathcal{J}_1\sin\theta_\mu\sin\theta_\nu\cos(\phi_\mu-\phi_\nu) \nonumber\\& 
-\mathcal{J}_2\sin\theta_\mu\sin\theta_\nu\cos(\phi_\mu+\phi_\nu) \big],
\label{mf} 
\end{align}
where $h_\mu=h\cos\theta_\mu$; $\xi_{\mu\nu}=1$ when $\mu$ and $\nu$ are nearest neighbours, and 0 otherwise (Fig.~\eqref{kagome}). We have taken the magnetic field, $h$, to be of order $S$.
$\mathcal{N}=3N$ is the number sites and $N$ is the number of unit cells. 
For $\mathcal{J}_2=0$, the classical energy is clearly minimized when $\phi_{\mu}$ is a constant, independent of the sublattice index.
Once $\mathcal{J}_2\neq0$, the energy is minimized for $\phi_{\mu}=0$ or $\pi/2$, depending on the sign of $\mathcal{J}_2$.  However,
since the sign of $\mathcal{J}_2$ can be changed by the aforementioned canonical transformation, we will fix it to be positive henceforth.
In this case, the groundstate corresponds to $\phi_{\mu} = 0$.


 
 In the case when $h=\mathcal{J}_1 = 0$, the classical model exhibits an accidental degeneracy.\cite{kle0, kle, kle1}
 Indeed, introducing
parameters $\delta_1$ and $\delta_2$ such that
$\theta_B=\delta_1-\delta_2$, and $\theta_C=\delta_1+\delta_2$, we have,
 \begin{align}
 &\cos \delta_1= \frac{-\lambda\cos\theta_A}{\sqrt{1-\lb 1-\lambda^2\rb\cos^2\theta_A}},\\&\cos \delta_2= -\frac{\cos \delta_1}{(1-\lambda)\cos\theta_A}, \quad \lambda\neq 1;
 \label{onep}
 \end{align}
where $\lambda=\mathcal{J}_2/\mathcal{J}_z$, and $\theta_A$ is arbitrary. 

 

These equations are identical to those that arise in the classical limit of the XXZ model, studied previously in Ref.~[\onlinecite{kle0, kle, kle1}].
It follows that including a nonzero $h$ field, 
the topology of the classical phase diagram is the same as the XXZ model, but the interpretation of the phases is different.
For example, the superfluid phase in the XXZ model is replaced by the CFM phase in our case ($\mathcal{J}_1=0$), which
only has a broken $Z_2$ symmetry, in contrast to the broken U(1) of the superfluid.

At $h_N^c=\mathcal{J}_2 + 2\mathcal{J}_z -\sqrt{4\mathcal{J}_z\lb\mathcal{J}_z-\mathcal{J}_2\rb-7\mathcal{J}_{2}^2 }$,
a phase transition occurs between the CFM phase and a phase with no CFM order, but with a finite magnetization $m \approx \pm 1/6$ 
in the $z$-direction.
In contrast to the XXZ model,\cite{kle1,kedar}
the total $S_z$ is not conserved here, hence these lobes retain a small finite magnetic susceptibility.\cite{juan}
Also, note that there exists  another classical phase boundary between the CFM and the FP state at $h_{F}^c=4(\mathcal{J}_z+\mathcal{J}_{2})$.
  

\section{Linear spin-wave analysis}  

In the previous section, we uncovered the classical phase diagram of the $Z_2$-invariant quantum kagome ice model.  In this section we perform a linear spin wave analysis of this model to study the role of quantum fluctuations.\cite{kedar, kle,kle1,kle0}  Our main goal is to investigate the role of the quantum fluctuations introduced by the $\mathcal{J}_2$ term
in lifting the extensive degeneracy of the classical ground state.
To facilitate the spin wave expansion, we rotate the coordinate axes so that the $z$ axis coincides with the local direction of the 
classical polarization.
We rotate by the angle $\theta_l$ about the $y$ axis, which corresponds to the rotation
matrix,
\begin{align}
 \mathcal{R}_y(\theta_l)= \begin{pmatrix}
  \cos\theta_l& 0& \sin\theta_l \\
0& 1&0 \\
-\sin\theta_l & 0& \cos\theta_l \\  
 \end{pmatrix}.
 \label{rot}
\end{align}
%
 Following the standard approach,\cite{jon}  we employ a three-sublattice Holstein Primakoff transform with the bosonic operators, $b^\dg_{l\mu}, b_{l\mu}$. After Fourier transform we obtain
\begin{align}
&H=\mathcal{E}_c+ S\sum_{\bo,\mu,\nu}\lb \mathcal{C}_{\mu\nu}^0\delta_{\mu\nu} +\mathcal{C}_{\mu\nu}^-\rb \lb b_{\bo \mu}^\dagger b_{\bo \nu}+b_{-\bo \mu}^\dagger b_{-\bo \nu}\rb\label{main}\\&\nonumber +\mathcal{C}_{\mu\nu}^+ \lb b_{\bo \mu}^\dagger b_{-\bo \nu}^\dagger +b_{-\bo \mu} b_{\bo \nu}\rb,
\end{align}
where $\mu,\nu=A,B,C$ label the sublattices, and the coefficients are given by \bea
\boldsymbol{\mathcal{C}}^0=\text{diag}\lb\chi_{AA},\chi_{BB},\chi_{CC}\rb,
\eea

\begin{align}
& \boldsymbol{\mathcal{C}}^\pm= \begin{pmatrix}
  0& \Delta_{AB}^\pm\gamma_{AB}&\Delta_{CA}^\pm\gamma_{CA}^* \\
\Delta_{AB}^\pm\gamma_{AB}^*& 0& \Delta_{BC}^\pm\gamma_{BC}\\
 \Delta_{CA}^\pm\gamma_{CA}& \Delta_{BC}^\pm\gamma_{BC}^* & 0 \\  
 \end{pmatrix}.
 \label{cmat}
\end{align} 
For the kagome lattice, we have $ \gamma_{AB}=\cos k_1, ~\gamma_{CA}=\cos k_2, ~\gamma_{BC}=\cos k_3,$
with $ k_p=\bo\cdot \bold{a}_p;$~ $ p=1,2,3$, where
\begin{align}
&\chi_{\mu\mu}=\frac{h_\mu-\sum_\nu \xi_{\mu\nu}\Delta_{\mu,\nu}^{zz}}{2};\quad \Delta_{\mu,\nu}^{\pm}=\frac{\lb \Delta_{\mu,\nu}^{xx}\pm \Delta_{\mu,\nu}^{yy}\rb}{2},
\end{align}
and
\begin{align}
 &\Delta_{\mu,\nu}^{zz}=\mathcal{J}_z\cos\theta_\mu\cos\theta_{\nu}-\mathcal{J}_{+}\sin\theta_\mu\sin\theta_{\nu},\nonumber\\&  \Delta_{\mu,\nu}^{xx}=\mathcal{J}_z\sin\theta_\mu\sin\theta_{\nu}-\mathcal{J}_{+}\cos\theta_\mu\cos\theta_{\nu},\nonumber\\&  \Delta_{\mu,\nu}^{yy}=\mathcal{J}_{-}; \quad h_\mu= h\cos\theta_\mu,
\end{align}
where $\mathcal{J}_\pm = \mathcal{J}_1 \pm \mathcal{J}_2$.

It is convenient to write Eq.~\eqref{main} in terms of the Nambu operators  $\Psi^\dg_\bo= (\psi^\dg_\bo, \thinspace \psi_{-\bo} )$  and $\psi^\dg_\bo=(b_{\bo A}^{\dg}\thinspace b_{\bo B}^{\dg}\thinspace b_{\bo C}^{\dg})$:
\begin{align}
&H_{XYZ}={\mathcal{E}}+ S\sum_{\bo}\Psi^\dg_\bo \mathcal{H}(\bo)\Psi_\bo,
\label{hp}
\end{align}
   with $\mathcal{E}=\mathcal{E}_c-\mathcal{E}_{lo}$; ~$\mathcal{E}_{lo}={\mathcal{N}}{S}\lb\chi_{AA}+\chi_{BB}+\chi_{CC}\rb/3 $ and  
\begin{align}
&\mathcal{H}(\bo) = \sigma_0\otimes\lb \boldsymbol{\mathcal{C}}_0 +\boldsymbol{\mathcal{C}}_- \rb + \sigma_x\otimes \boldsymbol{\mathcal{C}}_{+},
\end{align} 
where $\sigma_0$ is a $2\times 2$ identity matrix ($\mathbf{I}_{2\times 2}$) and $\sigma_x$ is a Pauli  matrix. 
 The Hamiltonian (Eq.~\eqref{hp}) can be diagonalized by the generalized Bogoliubov transformation \cite{jean,jean1}
\begin{align}
\Psi(\bo)= \mathcal{P}Q(\bo),
\label{bogo}
\end{align}
where $\mathcal{P}$ is a $6\times 6$ matrix, and  $Q^\dg_\bo= (\mathcal{Q}_\bo^\dg,\thinspace \mathcal{Q}_{-\bo})$ with $ \mathcal{Q}_\bo^\dg=(\alpha_{\bo A}^{\dg}\thinspace \alpha_{\bo B}^{\dg}\thinspace \alpha_{\bo C}^{\dg})$ being the quasiparticle operators. The matrix $\mathcal{P}$ satisfies the relations,
\begin{align}
&\mathcal{P}^\dg \mathcal{H} \mathcal{P}= \epsilon(\bo); \quad \mathcal{P}^\dg \eta \mathcal{P}= \eta,
\label{eig}
\end{align}
with
$\eta=
\text{diag}(
 \mathbf{I}_{3\times 3}, -\mathbf{I}_{3\times 3} )$ and $\epsilon(\bo)$ being the diagonal matrix of the quasiparticle energy eigenvalues.
 This is equivalent to saying that we need to diagonalize a
 matrix $\eta\mathcal{H}$, whose eigenvalues are given by $\eta\epsilon(\bo)=[\epsilon_\mu(\bo), -\epsilon_\mu(\bo)]$ and the columns of $\mathcal P$ are the corresponding eigenvectors. From Eq.~\eqref{eig} we have $\mathcal{P}^\dg =\eta \mathcal{P}^{-1}\eta$ and $\mathcal{P}^{-1}\eta \mathcal{H}(\bo)\mathcal{P}=\eta\epsilon(\bo)$ yielding 
\begin{align}
H_{XYZ}=\mathcal{E}_g +S\sum_{\bo\mu}  \epsilon_\mu(\bo)\lb\alpha_{\bo\mu}^\dg\alpha_{\bo\mu} +\alpha_{-\bo\mu}^\dg\alpha_{-\bo\mu}\rb,
\end{align}
with the ground state energy given by
\begin{align}
\mathcal{E}_g=\mathcal{E}_c-\mathcal{E}_{lo}+S\sum_{\bo,\mu}\epsilon_\mu(\bo).
\label{gr}
\end{align}

\begin{figure}[t]
\centering
\includegraphics[width=3.5in]{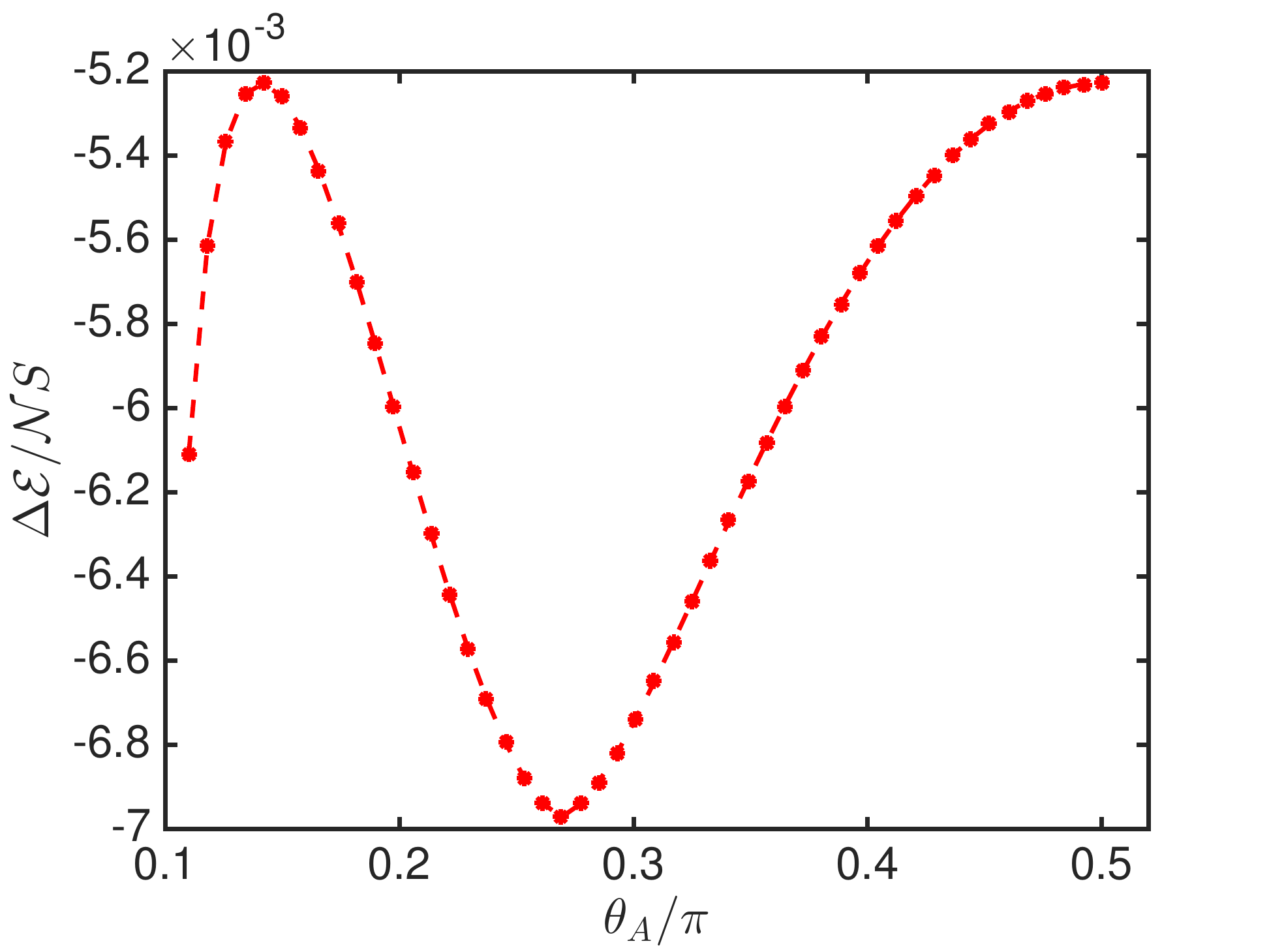}
\caption{Color online. The zero point quantum correction to the classical ground state as a function of $\theta_A$ with $\mathcal{J}_2=0.3$; $h=\mathcal{J}_1=0$; and $\mathcal{J}_z=1$.}
\label{Ground_corr}
\end{figure}
\subsection{Quantum selection of the ground state}  
At $h=0$, the classical ground state is highly degenerate.  
To see how the $\mathcal{J}_2$ term lifts this degeneracy, we analyze an expression for the quantum fluctuation correction to the classical energy, which is given by,
\begin{align}
\Delta \mathcal{E} =-\mathcal{E}_{lo}+S\sum_{\bo,\mu}\epsilon_\mu(\bo).
\end{align}
\begin{figure}[ht]
\centering
\includegraphics[width=3.5in]{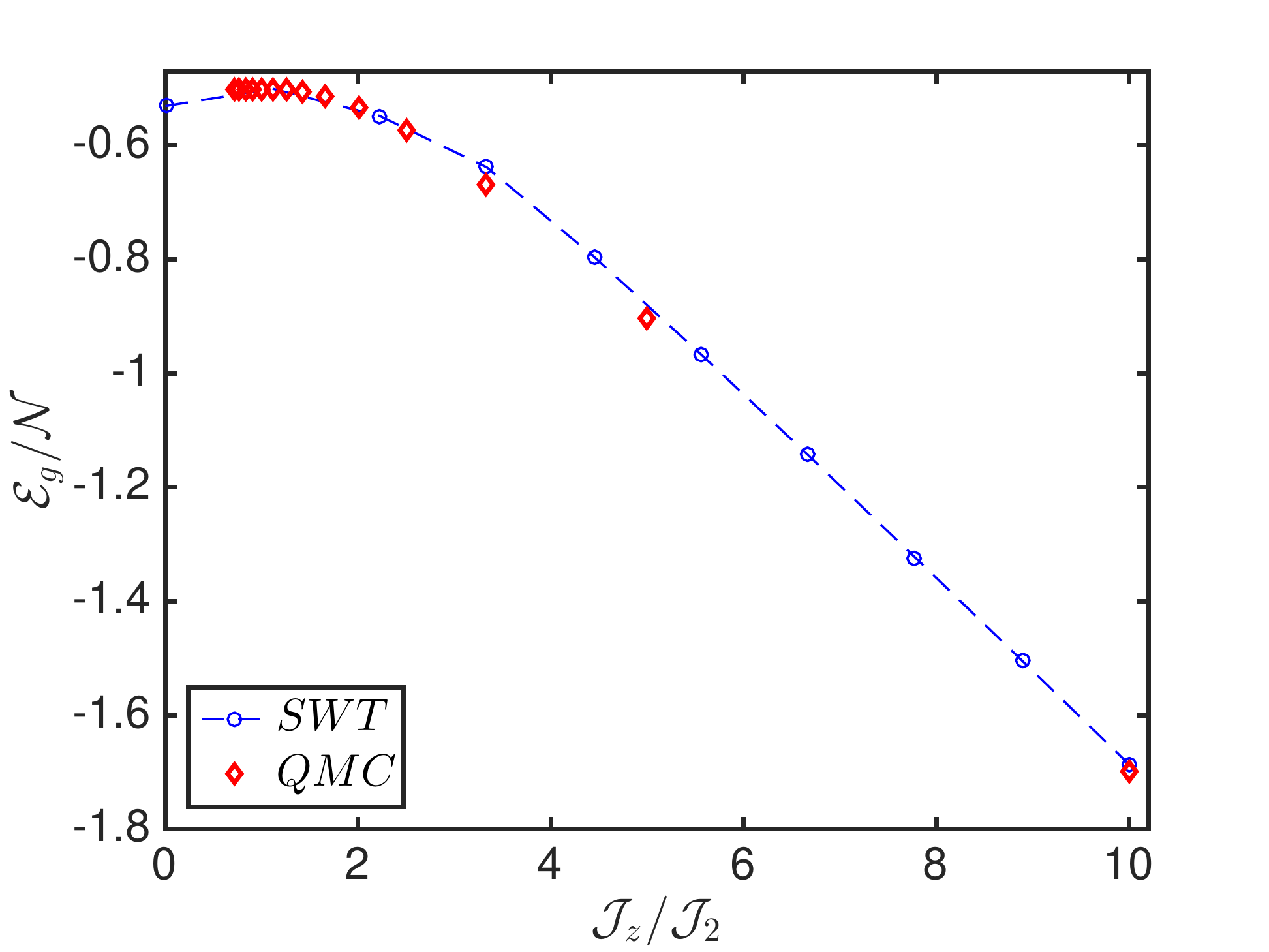}
\caption{Color online. Ground state energy per site vs. $\mathcal{J}_z/\mathcal{J}_2$ on kagome lattice for $S=1/2$ at $h=\mathcal{J}_1=0$. The QMC result \cite{juan1}  is with system size $V=3\times L\times L$ and $L=24$ at inverse temperature $\beta=96$.}
\label{true_gr}
\end{figure}
Figure \eqref{Ground_corr} shows the plot of $\Delta \mathcal{E}$ as a function of $\theta_A$, which parameterizes the different classical ground states. Quantum fluctuations select a particular $\theta_A$ as the lowest energy state.  This value of $\theta_A$ is a 
solution of the equation $\cos^2\theta_A\approx(1-2\lambda)/(1-\lambda^2)$.  
\begin{figure}[ht]
\centering
\includegraphics[width=3.5in]{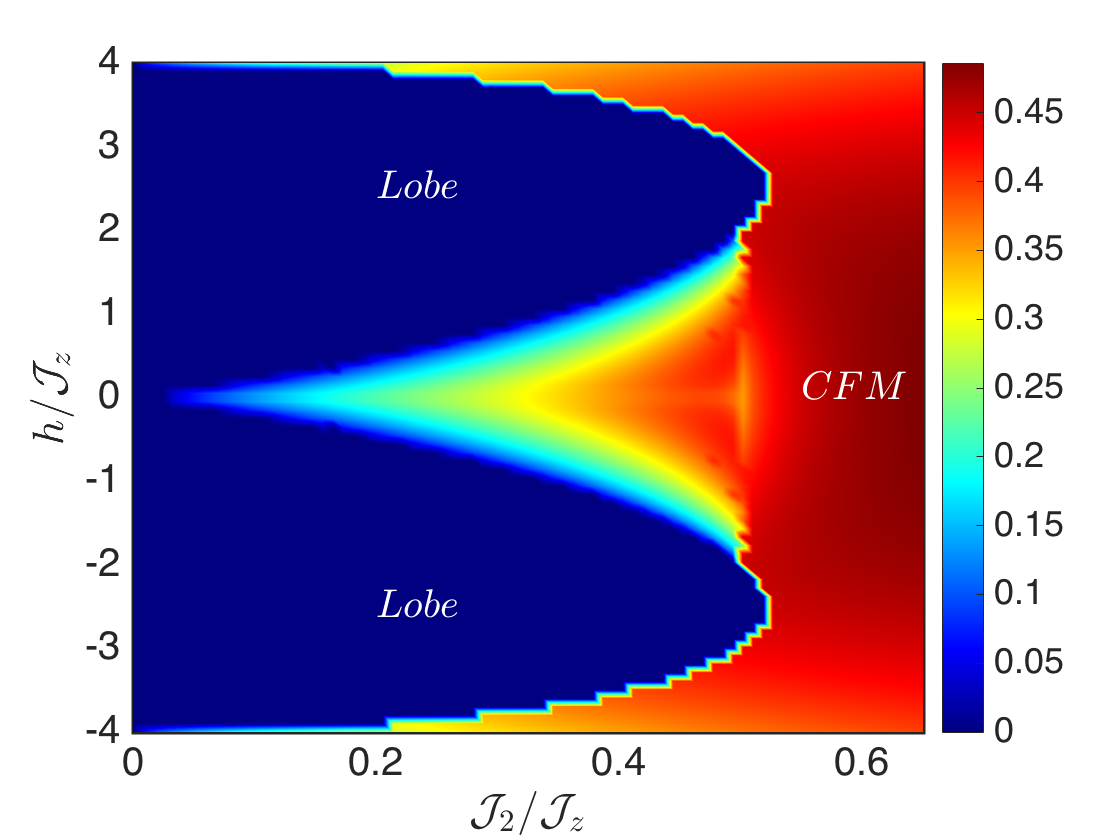}
\caption{Color online. The spin wave ground state phase diagram, as a function of $h/\mathcal{J}_z$ and $\mathcal{J}_2/\mathcal{J}_z$ for  $\mathcal{J}_1=0$. The colormap represents $\la S_x\ra$ per site. 
}
\label{phase_lobe}
\end{figure}

\begin{figure}[ht]
\centering
\includegraphics[width=3.5in]{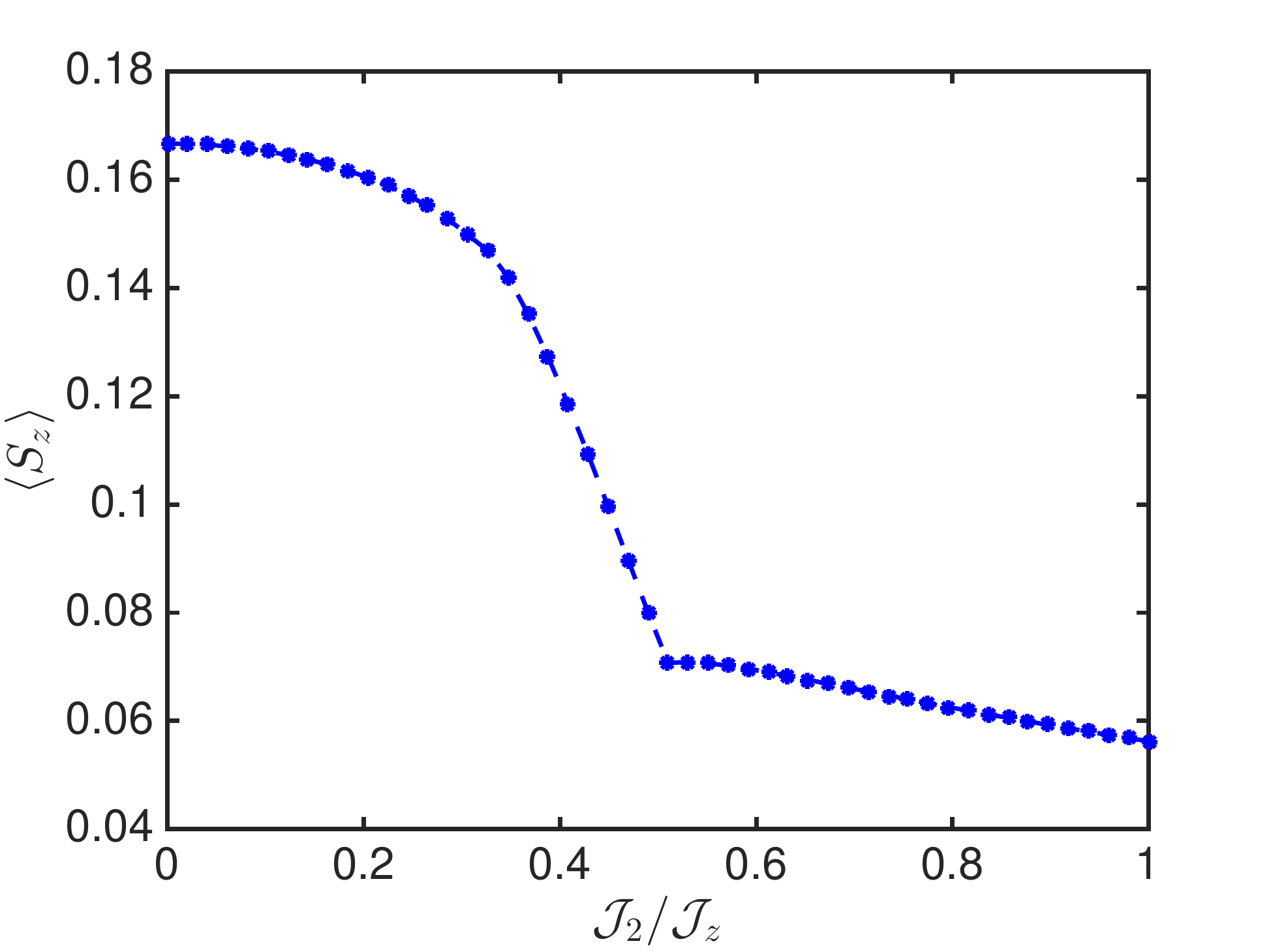}
\caption{Color online. $\langle S_z \rangle$  vs. $\mathcal{J}_2/\mathcal{J}_z$ entering the lobes at $h/\mathcal{J}_z=0.9$. {The trend captured in QMC simulation \cite{juan} is consistent with this spin wave calculation}. }
\label{Sz_lobe}
\end{figure}
\begin{figure}[ht]
\centering
\includegraphics[width=3.5in]{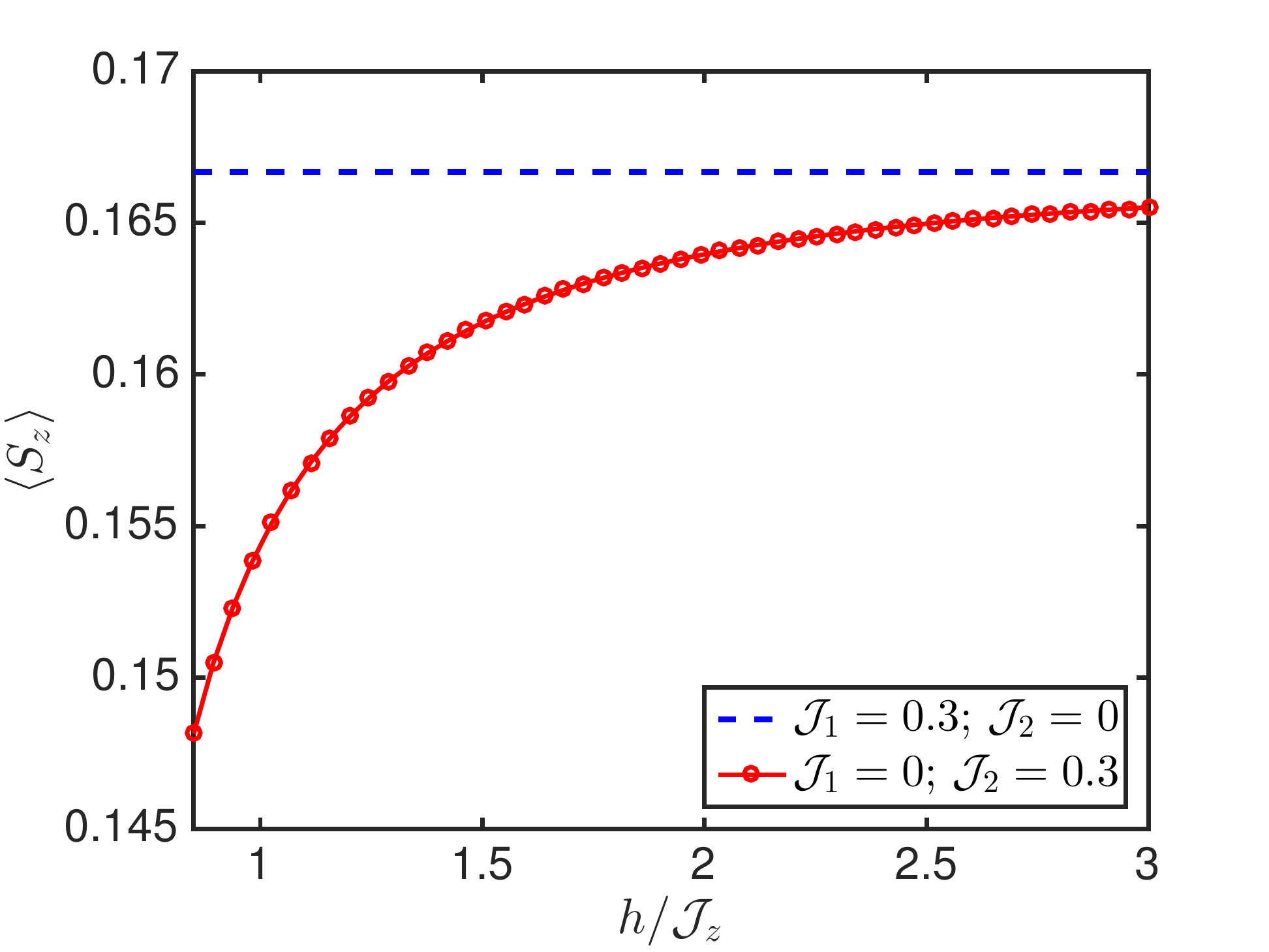}
\caption{Color online. $\langle S_z \rangle$ vs. $h/\mathcal{J}_z$ entering the lobes.  The dash line corresponds to the U(1) model, which is well-known to be the same in QMC simulation. The symbol shows the trends of  the $Z_2$ model. There is no corresponding QMC result reported in  Ref.~[\onlinecite{juan}]. }
\label{Sz_h}
\end{figure}
In Fig.~\eqref{true_gr} we show the ground state energy as a function of $\mathcal{J}_z$, together with the exact energy obtained by unbiased QMC simulations.\cite{juan1} Clearly, the linear SWT captures the trend of the dependence of the ground state energy on $\mathcal{J}_z/\mathcal{J}_2$. 
{At the rotationally symmetric point $\mathcal{J}_z=\mathcal{J}_2=1$ and $\mathcal{J}_1=h=0$, Eq.~\eqref{k1} becomes
\begin{align}
H&=\sum_{\la lm\ra}[S_{l}^zS_{m}^z+\frac{1}{2}\lb S_{l}^+S_{m}^+ + S_{l}^-S_{m}^-\rb],\nonumber\\&=\sum_{\la lm\ra}[-S_{l}^xS_{m}^x+S_{l}^yS_{m}^y+S_{l}^zS_{m}^z].
\end{align}
In the $S_x$-quantization axis, $S_\pm= S_z\pm iS_y$, hence the classical ground state is exact,  which gives $\mathcal{E}_g=\mathcal{E}_c=-0.5$ and corresponds to the maximum in Fig.~\eqref{true_gr}.}

\subsection{Quantum corrections to the magnetization}

Next, we examine the $h \neq 0$ region of the phase diagram.  {In the purely classical limit, $\mathcal{J}_{1,2}=0$ the addition of a symmetry-breaking field provides an 
additional constraint.  Namely, for $h=0$, each triangular plaquette of the lattice realizes a configuration with
either two $S^z = + 1/2$, one $S^z = - 1/2$, or one $S^z = + 1/2$, two $S^z = - 1/2$.  In contrast, for
$h>0$ say, only the two $S^z = + 1/2$, one $S^z = - 1/2$ configuration is energetically allowed for each triangle
(resulting in a magnetization per spin of $m= 1/6$).
Thus, the degeneracy is reduced compared to
the $h=0$ case, but still remains extensive (up to the critical field $h_{F}^c$). 
Including linear spin-wave corrections,}
we first compute the expectation value of the sublattice magnetization,  given by\cite{jean1, kle1}
\begin{align}
\la \bold{S}_\mu\ra =(SN-\sum_\bo \sum_{n=4}^6 |\mathcal{P}_{\ell,n}|^2)\bold{n}_{\mu},
\end{align}
where $\ell= 1, 2, 3$ for the sublattice $\mu=A,B,C$ respectively.  Note that the expectation value of $S_x$ plays a role analogous to the condensate fraction in Ref.~\onlinecite{juan}.  This will be useful in comparing our results to the QMC results of Ref.~\onlinecite{juan}.

{ In Fig.~\ref{phase_lobe}, we show a plot of the expectation value of $S_x$ as a function of $h/\mathcal{J}_z$ and $\mathcal{J}_2/\mathcal{J}_z$.  The phase diagram consists of three distinct phases. There is a CFM phase with $\la S_x\ra\neq 0$, and an FP phase which appears for $h/\mathcal{J}_z>4$ (not shown).   In addition, our theory captures the appearance of two $m\approx \pm 1/6$ lobes for $\mathcal{J}_2/\mathcal{J}_z<0.5$ and $ h/\mathcal{J}_z\neq 0$, with vanishing $\la S_x\ra$ and $\la S^+_lS^+_m\ra\neq 0$. These lobes are somewhat analogous to the well-known fixed-magnetization lobes of the XXZ model,\cite{xu, isa, roger, kedar, senn}
however, an important difference is that a finite susceptibility is retained throughout the lobes as mentioned above. This is a consequence of the fact that the Hamiltonian only has $Z_2$ symmetry, instead of U(1) in the XXZ case. Another feature of the phase diagram is that linear spin wave theory captures the same transition point ($\mathcal{J}_{\pm\pm}/\mathcal{J}_z=0.5$)  as QMC simulations \cite{juan} from the  lobes to CFM. This is in contrast to the U(1) model,\cite{isa, kle1} which suggests that quantum fluctuations are small in the  model with $\mathcal {J_{\pm\pm}} \neq 0; \mathcal{J_{\pm}}=0$}.
{ In the QMC simulations of the $\mathcal{J}_1=0$ Hamiltonian,\cite{juan} the nature 
of the ground state within the lobes is clearly distinct from the XXZ model,\cite{isa, kle1}where a Valence Bond Solid (VBS) is observed. In contrast, Ref.~\onlinecite{juan} observes a featureless spin liquid state. This is despite the fact that a perturbative argument taking $\mathcal{J}_2/\mathcal{J}_z \ll 1$ suggests a similar VBS phase as in the XXZ case.\cite{ALMH} QMC simulation also captures a  finite susceptibility in the lobes as corroborated here with spin wave theory calculation. }
\begin{figure}[ht]
\centering
\includegraphics[width=3.5in]{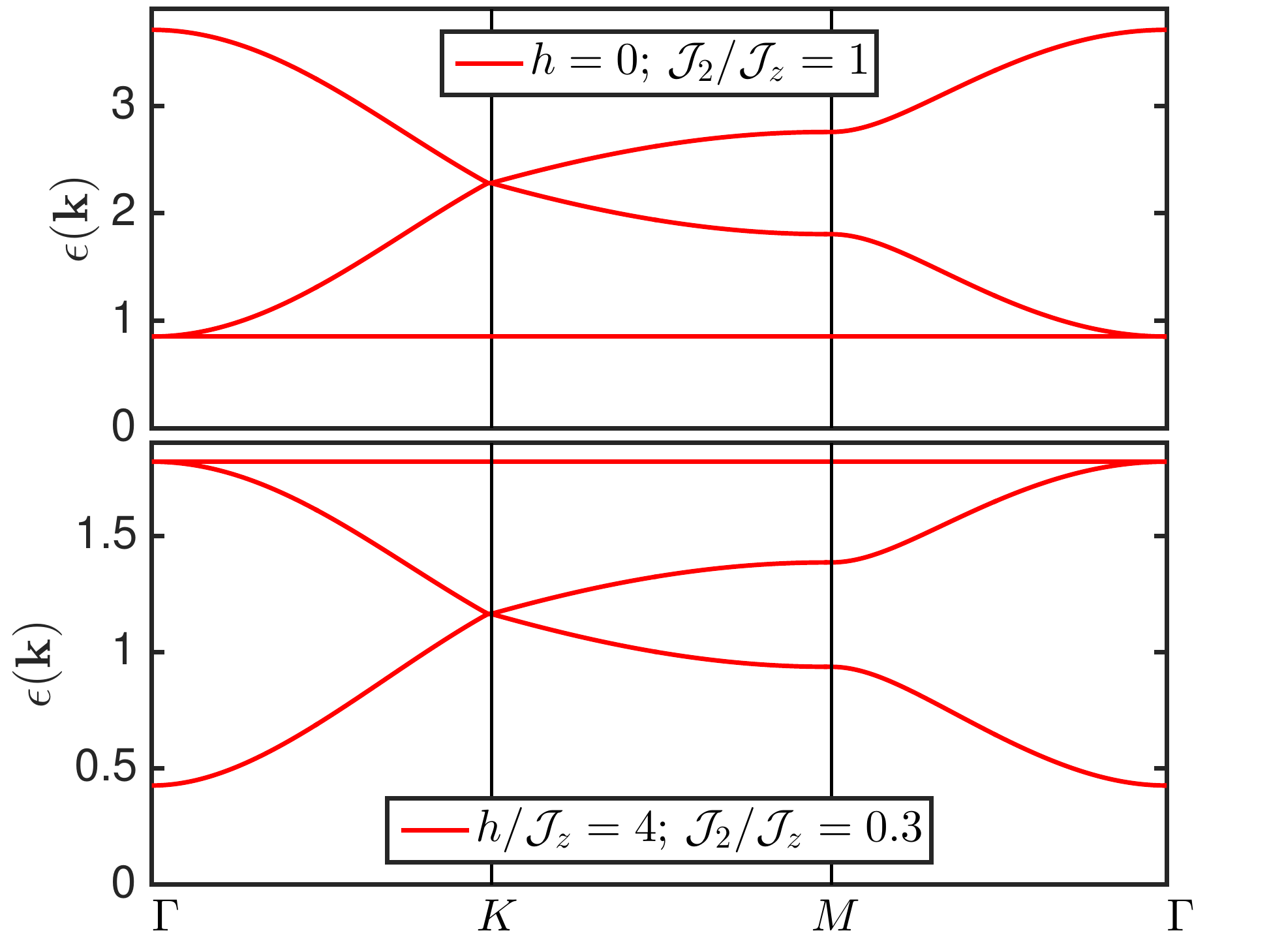}
\caption{Color online. The spin wave dispersion in units of $\mathcal{J}_z$  for the in-plane FM and CFM ordered states along the Brillouin zone paths with $\mathcal{J}_1=0$.  }
\label{pure_Z2_kag}
\end{figure} 
\begin{figure}[ht]
\centering
\includegraphics[width=3.5in]{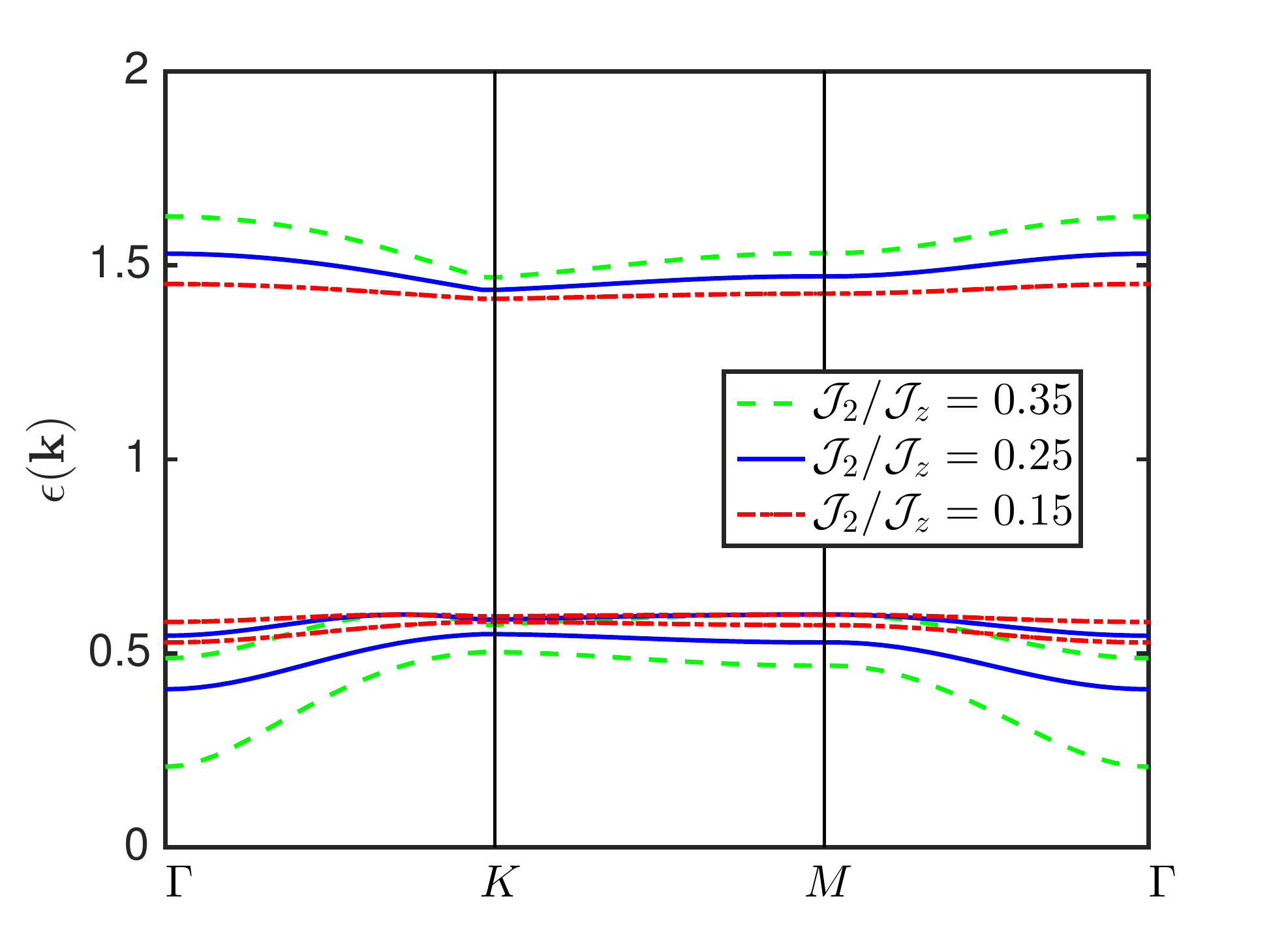}
\caption{Color online. The spin wave dispersion in units of $\mathcal{J}_z$  inside the $m \approx \pm 1/6$ lobe for $ \mathcal{J}_1=0$, $h/\mathcal{J}_z=1.2$.}
\label{Z2_lobe}
\end{figure}
  
In Fig.~\eqref{Sz_lobe}  we plot $\langle S_z \rangle$ entering the lobe at $h/\mathcal{J}_z=0.9$ and  Fig.~\eqref{Sz_h} shows the plot of  $\langle S_z \rangle$ as a function of the magnetic field inside the lobe. In the XXZ model, $\langle S_z \rangle = \pm 1/6$ inside the lobe.  
In our case, it is clear that we retain a small but finite susceptibility inside the lobe which is a consequence of the $Z_2$ symmetry of our model as discussed above.   These results are consistent with QMC simulations.\cite{juan} 

\subsection{Spin wave dispersion}   

Finally,  we examine the spin wave dispersion in the $m\approx \pm1/6$ lobes.
As shown in Fig.~\eqref{pure_Z2_kag},  a special feature of the  $\mathcal{J}_1 = 0$ model in zero field is that the spin wave spectrum does not have a soft mode, 
 which is a consequence of the discrete $Z_2$ symmetry. 
 There are two dispersive modes and one flat mode; the dispersive modes touch at the $K$ point (see Fig.~\ref{kagome}) in the Brillouin zone and the flat band always has the lowest energy.    The existence of the flat band is a well-known feature of the kagome lattice that follows from its geometry.\cite{kle1} When the field reaches the critical value $h=h_{N}^c$, one of the modes softens signifying the transition to the lobes.
In the lobe itself, the gap reopens and there are no soft modes again (see Fig.~\ref{Z2_lobe}).

\section{Conclusion}
Motivated by recent quantum Monte Carlo (QMC) simulations of the quantum kagome ice model,\cite{juan} we have presented another perspective of the ground state phase diagram of the model using the large-$S$ expansion. We focused mainly on the  $ \mathcal{J}_1=0$ model, which has not been studied before using this approach.   Exploring the parameter regimes of this model, we uncovered three distinct phases, 
including a canted ferromagnet (CFM), a fully-polarized (FP) state, and most interestingly finite-magnetization $m\approx\pm1/6$ ``lobes''.
This picture is consistent with the phases observed in QMC simulations.
At finite magnetic field and dominant Ising coupling, the nature of the $m\approx\pm1/6$ magnetized lobes is explicitly investigated and we find that all of the observables qualitatively agree with the QMC simulations.   
In particular, no evidence of any out-of-plane ($S_z$) spin order, which would be expected in the most likely candidate for valence-bond solid (VBS) order is found
within the lobes.  However, in order to more thoroughly address the possibility of VBS order within the lobes, it would be prudent to undertake a study using Schwinger bosons in the future.
 
\section{Acknowledgments}
The authors would like to thank Juan Carrasquilla for invaluable discussions, and also for providing us with details of his QMC simulations. 
Support was provided by NSERC, the Canada Research Chair program, and the Perimeter Institute (PI) for Theoretical Physics.
Research at Perimeter Institute is supported by the Government of Canada through Industry Canada and by the Province of Ontario through the Ministry of Research
and Innovation.

\end{document}